\documentstyle[12pt]{article}

\begin{document}
\thispagestyle{empty}

\begin{center}
{\large \bf Pseudogap phase formation in the crossover from\\
Bose-Einstein condensation to BCS superconductivity\\
in low dimensional systems}
\end{center}

\begin{center}
Valery~P.~Gusynin$^{1}$,
        Vadim~M.~Loktev$^{1}$,
                     Rachel~M.~Quick$^{2}$
 and \underline{Sergei~G.~Sharapov}$^{2, \dagger}$\\
$^{1}${\sl Bogolyubov Institute for
               Theoretical Physics,}\\
         {\sl  252143 Kiev, Ukraine} \\
$^{2}${\sl Department of Physics,
               University of Pretoria,}\\
            {\sl 0002 Pretoria, South Africa}
\end{center}

\noindent
A phase diagram for a 2D metal with variable carrier density
has been studied using the modulus-phase representation for the order
parameter in a fully microscopic treatment. This amounts to
splitting the degrees of freedom into neutral fermion and charged
boson degrees of freedom. Although true long range order is forbidden
in two dimensions, long range order for the neutral fermions is
possible since this does not violate any continuous symmetry. The
phase fluctuations associated with the charged degrees of freedom
destroy long range order in the full system as expected.
The presence of the neutral order parameter gives rise to new
features in the superconducting condensate formation in low
dimensional systems.

The resulting phase diagram contains a new phase which lies
above the superconducting (here Berezinskii-Kosterlitz-Thouless)
phase and below the normal (Fermi-liquid) phase. We identify this
phase with the pseudogap phase observed in underdoped high-$T_{c}$
superconducting compounds above their critical temperature. We also
find that the phase diagram persists even in the presence of weak
3-dimensionalisation. \\


\noindent
$^{\dagger}$
On leave from Bogolyubov Institute for Theoretical Physics
of the National Academy of Sciences of Ukraine, 252143 Kiev,
Ukraine \\

\noindent
Invited paper presented at First International Conference on\\
{\em New Theories, Discoveries, and Applications of Superconductors
and Related Materials}\\
February 19--24, 1998, Baton Rouge, Louisiana, USA

\newpage
The anomalous behavior of the normal state of high-temperature
superconductors (HTSC) \cite{Loktev.review}
(including the behavior of the spin susceptibility, resistivity,
specific heat and photo-emission spectra) has been recently interpreted
in terms of the formation of a pseudogap above the critical temperature,
$T_c$ \cite{Randeria}.

There are many approaches developed to explain this remarkable
phenomenon. However we present here the point of view that the
pseudogap phenomena is an inherent property of a low dimensional
system with a relatively small carrier density.

The formation of a pseudogap phase above $T_c$ has been explicitly
demonstrated in a very simple model non-relativistic 2D
Fermi-system \cite{Gusynin} and its extension to the quasi-2D
case \cite{Quick}.
The work is based on the peculiarities of the
Berezinskii-Kosterlitz-Thouless (BKT) phase formation which is
a two stage process. The only thing that one needs to analyze
the problem is to choose the appropriate variables.

    The simplest model Hamiltonian density for fermions
confined to a 2D volume $v$ reads
\begin{equation}
{\cal H} = \psi_{\sigma}^{\dagger}(x) \left(- \frac{\nabla^{2}}{2 m} - \mu
    \right) \psi_{\sigma}(x)
    - V \psi_{\uparrow}^{\dagger}(x)
       \psi_{\downarrow}^{\dagger}(x) \psi_{\downarrow}(x)
       \psi_{\uparrow}(x) \,, \label{Hamilton}
\end{equation}
where $x \equiv  \mbox{\bf r}, \tau$
denotes the position and imaginary time; $\psi_{\sigma}(x)$ is
a fermion field, $m$ is the effective fermion mass,
$\mu$ is a chemical potential,
$V$ is an effective local attraction constant; $\hbar = k_{B} = 1$.

The Hubbard-Stratonovich method is applied to write the statistical sum
$Z({v}, \mu, T)$ as a functional integral over fermi-fields (Nambu spinors)
and the auxiliary field $\Phi = V \psi_{\uparrow}^{\dagger}
\psi_{\downarrow}^{\dagger}$.
Contrary to the usual method for the calculation of $Z$ in $\Phi$,
$\Phi^{\ast}$ variables,
for a 2D system one must rewrite the order parameter $\Phi(x)$ in terms of
its modulus $\rho(x)$ and its phase $\theta(x)$ i.e. $\Phi(x) = \rho(x)
\exp[- i \theta(x)]$. This was originally stated by Witten in the context of
2D quantum field theory \cite{Witten}.  This replacement by modulus-phase
variables also demands the replacement
$\psi_{\sigma}(x) = \chi_{\sigma}(x) \exp{[i \theta(x)/2]}$.
From a physical point of view this entails replacing
the charged fermion particle $\psi_\sigma(x)$ by a neutral fermion
$\chi_\sigma(x)$ and a spinless charged boson $\exp{[i\theta(x)/2]}$.

As a result one obtains
\begin{equation}
Z(v, \mu, T) = \int \rho {\cal D} \rho {\cal D} \theta
\exp{[-\beta \Omega (v, \mu, T, \rho(x), \partial \theta (x))]} \,,
                 \label{partition.2D.effective}
\end{equation}
where
\begin{equation}
\beta \Omega (v, \mu, T, \rho (x), \partial \theta (x)) =
\frac{1}{V} \int_{0}^{\beta} d \tau \int d \mbox{\bf r}
\rho^{2}(x) - \mbox{Tr Ln} G^{-1}
                        \label{Effective.Action.2D}
\end{equation}
is  the one-loop effective action, which depends on the modulus-phase
variables. The action (\ref{Effective.Action.2D}) is expressed through
the Green function of the initial (charged) fermions written in the new
variables.

It is clearly impossible to obtain
$\Phi \equiv \langle \Phi(x) \rangle \ne 0$
at finite $T$ since this corresponds to the formation of homogeneous
long-range (superconducting) order which is forbidden by the
Coleman-Mermin-Wagner-Hohenberg theorem.  However it is possible to
obtain $\rho \equiv \langle \rho(x) \rangle \ne 0$ but at the same
time $\Phi = \rho \langle \exp[-i \theta(x)] \rangle = 0$
due to random fluctuations in the phase $\theta(x)$. We stress that
$\rho \ne 0$ does not imply long-range superconducting order
(which is destroyed by the phase fluctuations) and is therefore
not in contradiction with the above-mentioned theorem.

The effective action (\ref{Effective.Action.2D}) can be
approximately represented in the following form
\begin{equation}
\Omega (v, \mu, T, \rho, \partial \theta (x)) =
\frac{T}{2} \int_{0}^{\beta} \! \! \! d \tau
\int \! \! \! d \mbox{\bf r}
J(\mu, T, \rho(\mu,T)) (\nabla \theta)^{2} +
\Omega_{pot} (v, \mu, T, \rho).
                          \label{Effective.Action}
\end{equation}
The kinetic part of the action coincides with
the classical XY-model, where now however, $J$ depends on
$\mu$ and $\rho$. This allows to use the well-known equation for
the BKT temperature:
$T_{\rm BKT} = \frac{\pi}{2} J(\mu, T_{\rm BKT}, \rho(\mu,T_{\rm BKT}))$,
which should be solved together with the equations for
$\mu$ and $\rho$. These equations are obtained from
the effective potential $\Omega_{pot}(v, \mu, T, \rho)$.

Our investigation shows three regions in the 2D phase diagram \cite{Gusynin}.
The first is the superconducting (BKT) phase, where
$T < T_{\rm BKT}$ and $\rho \ne 0$. In this region there is algebraic order
and a power law decay of the superconducting correlations.
We note that for small carrier
densities (underdoped case) $T_{\rm BKT} = \epsilon_F/8$ where
$\epsilon_F$ is the Fermi energy. The second is the pseudogap phase
($T_{\rm BKT} < T < T_{\rho}$) where $\rho$ is still non-zero but
the correlations decay exponentially.  The temperature $T_{\rho}$ is that
at which $\rho$ becomes zero and it is interpreted as the temperature at
which the pseudogap opens. There is no phase transition at this
temperature.
The third is the normal (Fermi-liquid) phase
where $\rho = 0$. Note that $\Phi = 0$ everywhere. The unusual properties of
the second region, which lies between the superconducting and normal phases,
can be used to explain the pseudogap behavior in HTSC. For example in the
mean-field calculation of the paramagnetic susceptibility the parameter
$\rho$ plays the role of the energy gap $\Delta$ in the theory of convenient
superconductors.  Moreover, the full fermion Green function has a branch cut
and not a pole above $T_{\rm BKT}$ which is also in agreement with the
available data about the pseudogap.

It is important that all 2D results
obtained persist in the quasi-2D case. In this case one also has a phase
transition to a phase with long range order.  However the transition
temperature, $T_c$, always lies below the temperature $T_{\rm BKT}$ for the
quasi-2D system or coincides with $T_{\rm BKT}$ \cite{Quick}.
As one expects, in the high density limit,
all these temperatures tend asymptotically to the BCS temperature.

Note finally that
the description of the phase fluctuations and the BKT transition closely
resembles that given by Emery and Kivelson \cite{Emery}. However in their
phenomenological approach the field $\rho(x)$ does not appear while in the
present microscopic approach based on the modulus-phase
representation \cite{Witten} it
appears rather naturally.

\section*{Acknowledgements}
\noindent
We would like to thank Drs. E.V.~Gorbar, N.J.~Davidson, I.A.~Shovkovy
and V.M.~Turkowskii for numerous discussions.
R.M.Q and S.G.Sh also acknowledge the financial
support of the Foundation for Research Development, Pretoria.

\end{document}